\documentclass[twocolumn]{raa}     

\usepackage{graphicx, times}             
\usepackage{natbib}
\usepackage{amssymb,amsmath}
\usepackage{enumerate}
\bibpunct{(}{)}{;}{a}{}{,}
\usepackage{color}

\usepackage[pagebackref=true]{hyperref}

\def\kms{km s$^{-1}$}
\def\S{s$^{-1}$}
\def\mus{$\mu$s}
\def\HI{H\,{\small{I}} }
\def\s~{$\sim$}
\def\degree{$^{\circ}$}
\def\HIFAST{HiFAST}

\defcitealias{Jing2024}{Paper I}

\begin{document}

  \title{\HIFAST: An \HI Data Calibration and Imaging Pipeline for FAST
}
  \subtitle{III. Standing Wave Removal}

   \volnopage{Vol.0 (20xx) No.0, 000--000}      
   \setcounter{page}{1}          

   \author{Chen Xu \inst{1,3} 
   \and Jie Wang \inst{1,2,3}
   \and Yingjie Jing \inst{1} 
   \and Fujia Li \inst{4,5}
   \and Hengqian Gan \inst{1,6}
   \and Ziming Liu \inst{1,3}
   \and Tiantian Liang\inst{1,3}
   \and Qingze Chen \inst{1,3} 
   \and Zerui Liu \inst{1,3} 
   \and Zhipeng Hou \inst{1,3} 
   \and Hao Hu \inst{1,6}
   \and Huijie Hu \inst{3,1}
   \and Shijie Huang \inst{1,6}
   \and Peng Jiang \inst{1,7}
   \and Chuan-Peng Zhang \inst{1,7}
   \and Yan Zhu \inst{1,6} 
   }

   \institute{National Astronomical Observatories, Chinese Academy of Sciences,
   Beijing 100101, China; {\it xuchen@nao.cas.cn, jie.wang@nao.cas.cn, jyj@nao.cas.cn}\\
        \and
    Institute for Frontiers in Astronomy and Astrophysics, Beijing Normal University, Beijing 102206, China
        \and
    School of Astronomy and Space Science, University of Chinese Academy of Sciences, Beijing 100049, China \\
        \and 
    Department of Astronomy, University of Science and Technology of China, Hefei 230026, China \\
	\and 
    School of Astronomy and Space Science, University of Science and Technology of China, Hefei 230026, China
        \and
    CAS Key Laboratory of FAST, National Astronomical Observatories, Chinese Academy of Sciences, Beijing 100101, China
        \and
    Guizhou Radio Astronomical Observatory, Guizhou University, Guiyang 550000, China
    \\
\vs\no
   {\small Received 20xx month day; accepted 20xx month day}}

\abstract{The standing waves existed in radio telescope data are primarily due to reflections among the instruments, which significantly impact the spectrum quality of the Five-hundred-meter Aperture Spherical radio Telescope (FAST). Eliminating these standing waves for FAST is challenging given the constant changes in their phases and amplitudes. Over a ten-second period, the phases shift by 18\degree\ while the amplitudes fluctuate by 6 mK. Thus, we developed the fast Fourier transform (FFT) filter method to eliminate these standing waves for every individual spectrum. The FFT filter can decrease the root mean square (RMS) from 3.2 to 1.15 times the theoretical estimate. Compared to other methods such as sine fitting and running median, the FFT filter achieves a median RMS of approximately 1.2 times the theoretical expectation and the smallest scatter at 12\%. Additionally, the FFT filter method avoids the flux loss issue encountered with some other methods. The FFT is also efficient in detecting harmonic radio frequency interference (RFI). In the FAST data, we identified three distinct types of harmonic RFI, each with amplitudes exceeding 100 mK and intrinsic frequency periods of 8.1, 0.5, and 0.37 MHz, respectively. The FFT filter, proven as the most effective method, is integrated into the \HI data calibration and imaging pipeline for FAST (\HIFAST, \url{https://hifast.readthedocs.io}).
}

   \authorrunning{Xu C. et al.}            
   \titlerunning{Standing Wave Removal}  

   \maketitle
%
%
\section{Introduction}
Standing waves, sometimes also known as the baseline ripple or Fixed-pattern Noise \citep{Heiles2005}, are a prevalent issue for FAST and many other single-dish radio telescopes, such as the 305-m Arecibo telescope \citep{Briggs1997} and the 64-m Parkes telescope \citep{Barnes2001, Reynolds2017}.
They stem from the reflections of electromagnetic (EM) waves between two surfaces or endpoints, such as the feed cabin with the dish or the joints of optical fibers.
The EM wave could source from feed, ground, and strong radio sources in the sky.  The coherent superposition of these waves onto the target's signal results in sinusoidal standing waves, which cause the deterioration of the spectral baseline \citep{Liu2022}. 

Many attempts have been made to remove standing waves in radio telescopes. Some instruments and new designs have been developed on the hardware \citep[e.g.][]{Padman1977, Goldsmith1980}, while residual standing waves still exist in the telescope.
As for software-based approaches, fitting a sine function (e.g., software {\small{UniPOPS}}\footnote{{\small{UniPOPS}} document \url{https://www.gb.nrao.edu/~rmaddale/140ft/unipops/unipops_7.html}}) and running median/mean method (e.g., SoFiA 2\footnote{HI Source Finding Application 2 \url{https://github.com/SoFiA-Admin/SoFiA-2/}} \citep{Westmeier2022}) are widely used to mitigate the standing waves. 
However, as the parameter complexity in function fitting and negative flux shadow caused by running methods \citep{Barnes2001, Minchin2010}, these methods could not clear the residual standing wave completely. So \citet{Winkel2011} suggested that the FFT filtering algorithms might be a better substitute.

Fourier analysis has been widely used to investigate the standing waves \citep[e.g.,][]{Fisher2003, Heiles2003, Heiles2005, Mcintyre2013, Reynolds2017, Li2021}. It decomposes the waves into a series of sine functions, and then an FFT filter could be applied to mitigate these standing waves.
The FFT-based methods have been implemented in many telescopes and reduction software \citep[e.g.,][and {\small{CASA}}\footnote{ Common Astronomy Software Applications (CASA) document \url{https://casadocs.readthedocs.io/en/stable/notebooks/single_dish_calibration.html}}]{Briggs1997, Heiles2003, Butcher2016, Liu2022}.

As FAST has a unique structure with an active reflector and a feed cabin flexibly driven by cables \citep{Nan2011}, the phase and amplitude of standing waves are supposed to always vary.  
In contrast to previous methods, the FFT filter method does not require a long integration spectrum and can operate on a single spectrum. It does not result in signal loss or subtract standing waves based on the exact phase and amplitude of each spectrum. 
As a result, the FFT filter is considered the optimal choice to eliminate standing waves in our FAST \HI data reduction pipeline \HIFAST.\footnote{\HIFAST\ cookbook  \url{https://hifast.readthedocs.io}} \citep[][hereafter \citetalias{Jing2024}]{Jing2024}. 

\citetalias{Jing2024} has presented an overview of the data processing pipeline, \HIFAST. \HIFAST\ is a modular, flexible, and dedicated calibration and imaging pipeline for the \HI FAST data.  
The pipeline consists of noise diode calibration, baseline subtraction, RFI flagging, standing wave removal using the FFT filter, flux and gain-curve calibration \citep[][Paper II]{Liu2024}, Doppler correction, stray radiation correction (Chen et al. 2024, in preparation, Paper IV), and finally the gridding to produce data cubes. 
As a part of our series of HiFAST papers, in this paper, all related issues on standing wave and harmonic RFI that could be identified with the FFT scheme will be studied in detail.  
~\\ 

This study is divided into seven sections. 
In Section \ref{sec:obs}, we provide an introduction to the observations and data reduction. 
The basic information of the standing waves in FAST is presented in Section \ref{sec:features}. 
Three methods for removing these standing waves, namely sine-fitting, running median, and FFT filter, are described in Section \ref{sec:methods}. 
Among these methods, we focus on the FFT filter due to its superior removal effects compared to the other methods, as discussed in Section \ref{sec:result}. 
Furthermore, we discuss the harmonic RFI discovered at FAST, which exhibits regularity in Fourier space, in Section \ref{sec:rfi}. 
Finally, our conclusions and discussions are presented in Section \ref{sec:conclusion}.

\section{Observations}\label{sec:obs}

As we pointed out in our \citetalias{Jing2024}, the presence of harmonic RFI with a frequency of 8.1 MHz posed a significant challenge in accurately determining the sensitivity of the telescope. 
Fortunately, this issue has been completely resolved as of July 2021. Therefore, all the data utilized in this study are from observations conducted after this date.

In this study, scan mode data are utilized for most tests. 
For point source examinations, a drift scan region observed on August 8th and 9th in 2021 is utilized. 
Furthermore, data from our ongoing blind drift survey, which focuses on the Andromeda galaxy and spans three observational seasons (2021/10--2022/02, 2022/10--2023/01, and 2023/09--2024/01), are used to illustrate the features of standing waves and harmonic RFI. 
To demonstrate the variation of the standing wave in different observational modes, the study incorporates additional data from  MultibeamOTF (On-The-Fly mapping) and Tracking modes for the analysis.

The observation employs a wide-band spectral backend with a resolution of 7.63 kHz or 1.61 \kms (at $z=0$). 
The data is recorded in two dual linear polarisations, \texttt{XX} and \texttt{YY}. 
For most of our observations, the high-power noise diode, which operates at 10 Kelvin, is turned on for 2 seconds and off for 298 seconds in a 5-minute iteration, with a spectrum sampling rate of 1/0.5~\S. 
Due to the presence of severe RFI in the frequency range of 1150 MHz to 1290 MHz (caused by communication and navigation satellites) \citep{Zhang2022, Xi2022}, the frequency range of 1300 MHz to 1450 MHz is primarily used for data reduction.

\section{Properties of the Standing Waves}\label{sec:features}

\begin{figure}
    \includegraphics[width=0.5\textwidth, angle=0]{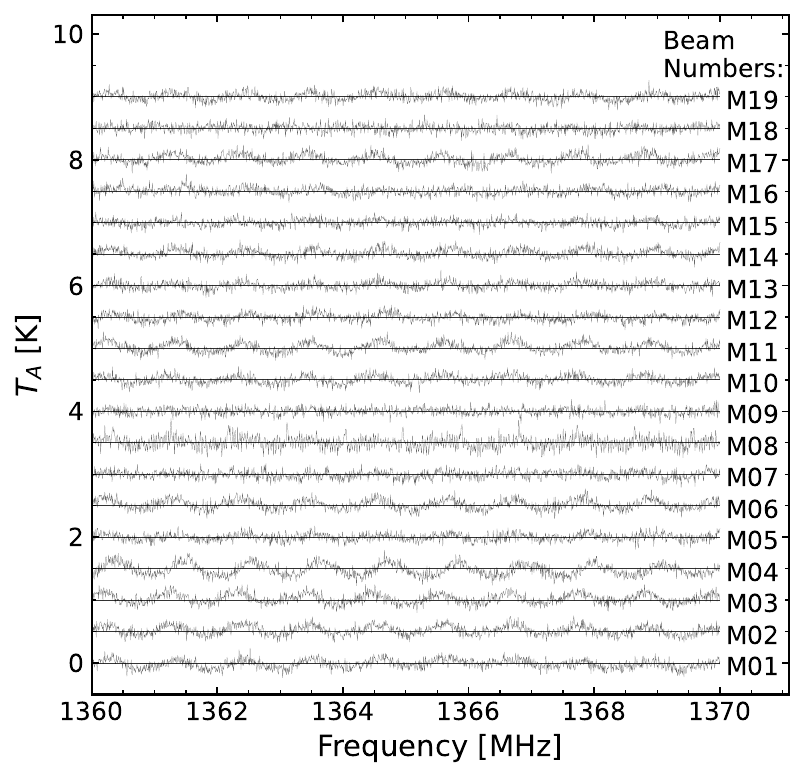}\centering
    \caption{The spectra of polar \texttt{YY} in 19 beams are displayed in real space. The beam numbers are indicated on the right side of the figure. To enhance visibility, each spectrum, which was averaged for 25 seconds, has been gradually offset by 0.5 K. The figure clearly illustrates that the amplitudes and phases of the standing waves vary across the beams.
    We note that the noise level of M08 is higher than others due to the 0.5-MHz harmonic RFI, as described comprehensively in Section~\ref{sec:0.5rfi}.
    }
    \label{fig:bms}
\end{figure}

Figure~\ref{fig:bms} illustrates the representation of standing waves in 19 beams with polarized \texttt{YY} by averaging the spectrum over a duration of 25 seconds. It is evident that these standing waves manifest as sinusoidal waves along the spectral axis. 
While the period of the sinusoidal waves appears to be consistent across different beams, the amplitudes and phases exhibit variations from one beam to another. 
In this section, we will present the analysis of the three quantities of sinusoidal waves: periods, phases, and amplitudes separately.

\subsection{Periods of the Standing Waves}\label{sec:period}

\begin{figure*}
    \includegraphics[width=\textwidth]{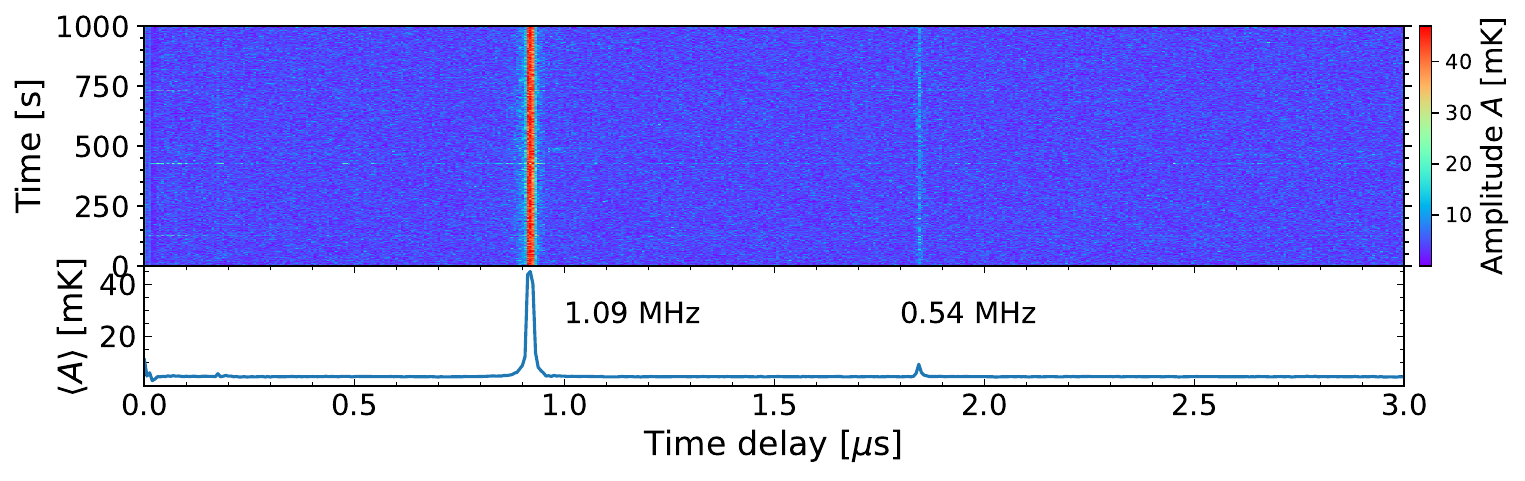}\centering

    \includegraphics[width=\textwidth]{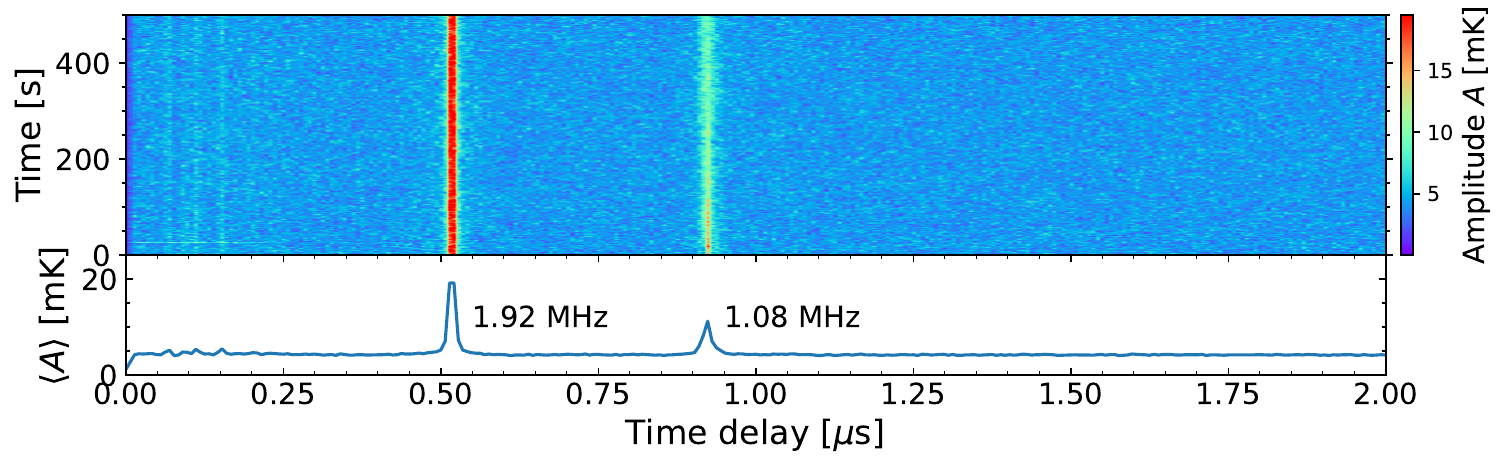}\centering
    
    \includegraphics[width=\textwidth]{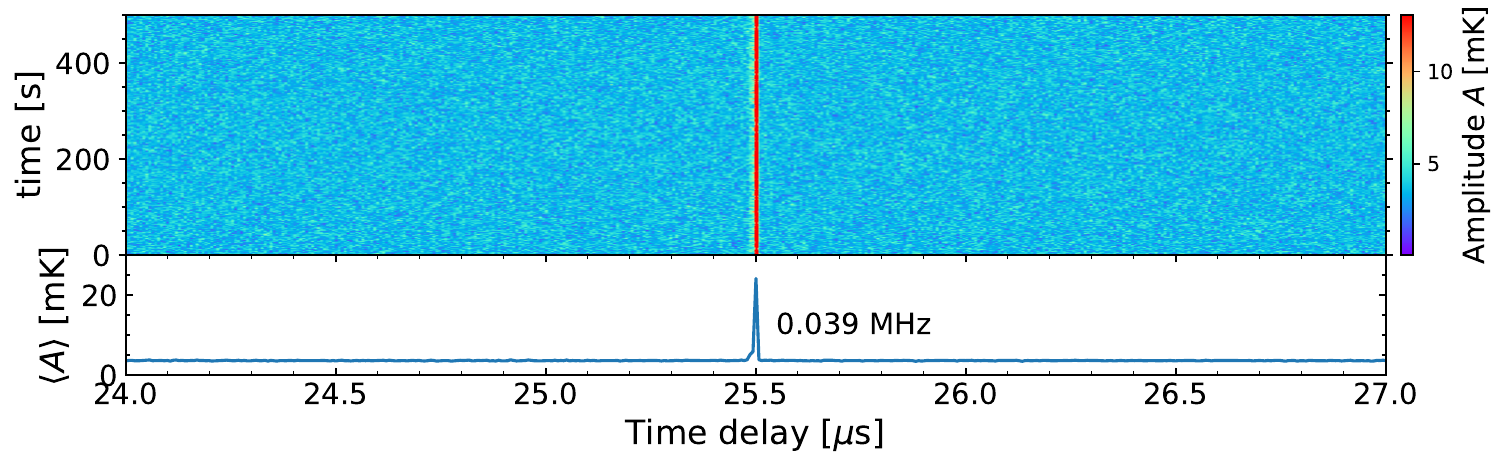}\centering

    \caption{Upper panel: The 1-MHz standing wave amplitudes in Fourier space. The spectrum below the image is the average of all the amplitudes along the time axis. Two vertical peaks stand for the 1.09-MHz (0.92 \mus) standing wave and its second harmonic wave, 0.54 MHz (1.84 \mus). Note that some faint horizontal lines come from the activated noise diode. 
    Middle panel: Similar to the upper panel, but the central peak indicates the 1.92-MHz (0.52 \mus) standing wave in the Fourier space. 
    Lower panel: The 0.039-MHz (25.5 \mus) standing wave in the Fourier space. 
    }
    \label{fig:fourier}
\end{figure*}

The presence of a standing wave arises from the wave's reflection between two surfaces or endpoints, such as the dish surface and receiver, or the two ends of an optical cable. The period or frequency interval $f$ of the standing wave can be determined using the following definition:
\begin{equation}
    f = \frac{c}{2L}, \label{equ:sdw}
\end{equation}
where $L$ is the radiation path length and $c$ is the speed of light. 
The primary standing wave of the FAST is reflected between the dish and the receiver cabin. In the case of this standing wave, the distance between the dish and receiver cabin is referred to as the focal length, denoted as $F$. According to the focal ratio ($F/D \approx 0.46$) provided by \citet{Jiang2019}, the distance between the receiver and the main reflector for FAST is approximately 138 m. This distance corresponds to a standing wave period of 1.087 MHz in frequency, which in turn corresponds to a velocity width of approximately 200 \kms at $L$ band. 
For comparison, other telescopes such as Arecibo and Parkes face similar issues, with their standing waves having periods of 1 MHz and 5.7 MHz, respectively \citep{Peek2011, Barnes2001}.
The upper panel of Figure~\ref{fig:fourier} shows the amplitudes of the 1.09-MHz standing wave and its second harmonic wave in Fourier space. 

To characterize standing waves in the Fourier space, we employ the time delay ($\tau$), which is the inverse of $f$. The definition of the time delay is as follows:

\begin{equation}
    \tau =\frac{1}{f} = \frac{2L}{c}.
    \label{equ:sdw-1}
\end{equation}

If we multiply a positive integer $N$ in Equation (\ref{equ:sdw-1}), it could describe the harmonic waves of the standing wave,
\begin{equation}
    \tau_N = N \frac{2L}{c}, N = 1, 2, \cdots.
\end{equation}
When $N=1$, it represents the fundamental frequency or time delay of the standing wave. In the case of the 1.09 MHz standing wave, the inverse of its period corresponds to a time delay of 0.92 microseconds. 
When $N=2$, it indicates a less intense second harmonic wave of the fundamental frequency, with a time delay of 1.84 microseconds in the frequency domain. The spectrogram displayed below illustrates the average of 1000 seconds' worth of spectra. 
It depicts two peaks denoting the 1.09 MHz (0.92 \mus) standing wave and its second harmonic wave at 0.54 MHz (1.84 \mus). Hereafter, we refer to the 1-MHz standing wave as the fundamental frequency of 1.09 MHz for brevity.

Normally, the fundamental frequency of the 1-MHz standing wave falls within the range of 1.07 to 1.09 MHz in the frequency domain. The central peak position, which represents the time delay of the standing wave, exhibits a slight fluctuation in the Fourier domain, typically within a range of $ \pm \Delta \tau$ due to noise and spectral leakage. 
This discrepancy is also influenced by the resolution in the discrete Fourier transform (DFT). The resolution ($\Delta \tau$) is reciprocally related to the frequency bandwidth ($W_{\rm{freq}}$), as expressed by the equation 
\begin{equation} 
\frac{\Delta \tau}{\mu{\rm{s}}} = \left( \frac{W_{\rm{freq}}}{\rm{MHz}} \right)^{-1}, \label{equ:dt} 
\end{equation}
indicating that a larger $W_{\rm{freq}}$ leads to a higher resolution. 
In the whole study, $W_{\rm{freq}}$ exceeds 60 MHz to empirically differentiate the components of the standing waves from other modes.

In addition to the prominent 1-MHz standing wave and its harmonic wave, additional standing waves have been identified in the FAST data. These two standing waves are also clearly visible in Fourier space, as depicted in the bottom two panels of Figure~\ref{fig:fourier}. 
One of these standing waves has a frequency of 1.92 MHz (corresponding to a period of 0.52 \mus) and is observed in certain beams, such as M03 \texttt{YY} and M06 \texttt{YY}. 
It is hypothesized that this standing wave may be a result of reflections occurring at a distance of 78 m, predominantly appearing in the data collected after October 2022 \footnote{In our data, we observed the presence of the 1.92-MHz standing wave in beams M02, M06 \texttt{XX}, M06, M09, M12, M16 \texttt{YY} from October 2022 to February 2023; and in beams M04, M06, M17 \texttt{XX}, M03, M04, M06, M16 \texttt{YY} from July 2023 to the end of 2023. This standing wave was not distinctly noticeable before October 2022, but still remains in 2024.}.

Another standing wave, which occurs at a frequency of 0.039 MHz (equivalent to 25.5 \mus), is relatively weak and is detected in about 19 beams. This wave is generated by reflections within the optical fiber, which spans a distance of roughly 3.8 km. 
Although it may not be readily apparent in an individual spectrum, it becomes noticeable in the Fourier spectrum displayed in the lower section of Figure~\ref{fig:fourier}, particularly in specific polarisations such as M08 \texttt{YY}. Its significance is more pronounced during extended periods of integration (e.g., surpassing 60 seconds) in Tracking or Snapshot modes.

~\\
Table~\ref{tab:sw} provides a summary of the three types of standing waves at FAST: 1.09-MHz with its second harmonic wave (0.54-MHz), 1.92-MHz, and 0.039-MHz standing wave.  The amplitude in this table corresponds to the intensity in Fourier space. 
We use the root mean square (RMS) in the five-minute integration and over the 5-MHz bandwidth to quantify the strength of standing waves in the unit of the theoretical RMS, which is defined as \citet{Jiang2020}: 
\begin{equation}
    \sigma_{\mathrm{theory}} = \frac{T_\mathrm{sys}}{\sqrt{n_\mathrm{p} \beta \Delta t}} \label{equ:theory_rms},
\end{equation} 
where $T_\mathrm{sys}$ is the systematic temperature in Kelvin, $n_\mathrm{p}$ is the number of polarisations utilised in the spectrum, $\beta$ equals 1.2 times channel resolution with the unit of Hz, and $\Delta t$ is the integration time in second.

\begin{table*}[ht]
	
    \setlength{\tabcolsep}{5pt}
    \renewcommand\arraystretch{1}
    \caption{Summary of the standing waves found in the FAST data.  
    }
    \begin{center}
    \begin{tabular}{ccccccc}
        \hline\noalign{\smallskip}
        Period / MHz  & Time delay / \mus & Amplitude / mK & RMS$^*$ / $\sigma_{\mathrm{theory}}$ & Weakened time & Weakened amplitude / mK & Comments\\
        \noalign{\smallskip}
        \hline
        \noalign{\smallskip}
        1.09 & 0.92  & \s~ 10 -- 50 & 3.2 & -          & -     &  $^a$ \\
        0.54 & 1.84  & \s~ 5 -- 10  & -    & 2022/03/02 & $<$ 5 & $^b$ \\
        1.92 & 0.52  & \s~ 7 -- 50  & 3.5  & -          & -     & $^c$ \\
        0.039 & 25.5 & \s~ 1 -- 20  & -    & 2022/09/27 & $<$ 5 & $^d$ \\
        \noalign{\smallskip}\hline
    \end{tabular}      
    \end{center}
{\small
$^*$ RMS in 5-MHz and 5-minute integration.

$^a$ The 1-MHz standing wave exists in every beam and varies. Only the RMS of M01, \texttt{YY} is measured in this table.

$^b$ This should be a harmonic component of the 1-MHz standing wave as it is closely coupled to the 1-MHz standing wave. Consequently, its RMS is challenging to measure independently. After the instrumental improvement, its Fourier amplitude is lower than the noise. 

$^c$ The 1.92-MHz standing wave was only observed in a limited number of polarisations and beams. The root mean square (RMS) value is determined solely through M03, \texttt{YY} (with an amplitude of approximately 50 mK), for instance. Refer to Section \ref{sec:period} for more information.

$^d$ It is extremely weak in most beams (and polarisations), so the Poisson noise and baseline flatness dominate in the RMS calculation. We do not give the RMS either. Also, its Fourier amplitude has decreased to the noise level after the instrumental improvement. See Section \ref{sec:period} for details. 
}
\label{tab:sw}
\end{table*}

Although it is impossible to remove the 1-MHz standing wave instrumentally, the deployment of fiber isolators in March 2022 successfully reduced its second harmonic wave. Similarly, the 0.039-MHz standing wave has been diminished since the replacement of fiber isolators and the cleaning of the fiber optic connectors in September 2022. The rest discussion will concentrate on the characteristics of the 1-MHz standing wave.

\subsection{Amplitudes and Phases of the Standing Waves}\label{sec:amp_phi}

\begin{figure}
    \includegraphics[width=0.5\textwidth, angle=0]{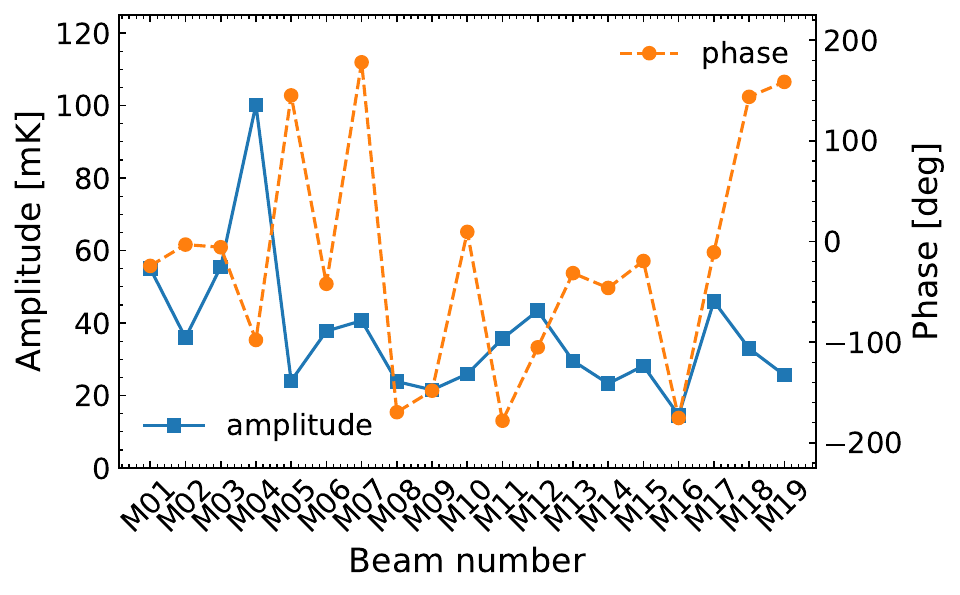}\centering
    \caption{Amplitudes (blue solid lines with square markers) and phases (orange dashed lines with circle markers) of the 1-MHz standing wave in 19 beams, polarisation \texttt{YY}. M04 has the highest amplitude, and the phases also change in beams.}
    \label{fig:bms_ap}
\end{figure}

In the Fourier domain, we identify the component with the greatest amplitude to measure the amplitudes and phases of the standing wave. Figure~\ref{fig:bms_ap} shows the 1-MHz standing wave's amplitudes and phases in FAST's 19 beams for polarisation \texttt{YY}, with M04 exhibiting the most obvious standing wave compared to the other beams. 
Moreover, the phases differ among various beams, a common occurrence for two polarisation channels and all beams in every observation data set. Therefore, it is necessary to analyse the data of each polarisation channel and beam individually.

Examining the variation in amplitudes and phases over an extended period is an interesting aspect to explore. 
To achieve this, we opted to utilize the spectral data from a sky range devoid of any detectable sources over eight days of M01, consisting of two MultibeamOTF scans, two Tracking observations, and four drift scans. 
The details for these observational data are summarized in Table~\ref{tab:obs}.

\begin{table*}[ht]

	\setlength{\tabcolsep}{8pt}
	\renewcommand\arraystretch{1}
	\caption{Observational information of the example data to explore the variation in amplitudes and phases over an extended period.
    The column of ZA indicates the average zenith angle during the observation.
    We note that the unit of the noise-diode periods ($t_\mathrm{delay}$, $t_\mathrm{Cal-On}$, and $t_\mathrm{Cal-Off}$) mentioned in this study is not precisely 1 second, but rather 1.00663296 seconds.
	}
    \begin{center}
	\begin{tabular}{ccccccc}
		\hline\noalign{\smallskip}
Obs date & Obs mode & Centre position & ZA / \degree &  $t_\mathrm{delay}$ / s  & $t_\mathrm{Cal-On}$ / s & $t_\mathrm{Cal-Off}$ / s\\
		\noalign{\smallskip}
        \hline
        \noalign{\smallskip}
2021/07/31 & MultibeamOTF & 01:33:57.11 +30:39:35.82 & 6.7 & 2 & 2 & 298 \\
2021/08/09 & Drift & 14:34:05.54 +00:49:29.75 & 24.9 & 2 & 2 & 298 \\
2022/02/12 & Drift & 00:43:06.26 +53:00:38.40 & 27.5 & 2 & 2 & 298 \\
2023/01/28 & Drift & 00:33:05.87 +52:59:50.50 & 27.5 & 2 & 2 & 298 \\
2023/07/22 & MultibeamOTF & 16:04:11.95 +43:39:36.86 & 20.9 & 2 & 2 & 298 \\
2023/07/24 & Tracking & 00:30:39.86 +37:24:13.61 & 16.6 & 1 & 1 & 15 \\
2023/07/29 & Tracking & 00:28:11.84 +36:36:57.08 & 16.7 & 1 & 1 & 15 \\
2024/02/14 & Drift & 00:33:05.57 +52:37:46.52 & 27.1 & 20 & 2 & 298 \\
		\noalign{\smallskip}\hline
	\end{tabular}      
    \end{center}
\label{tab:obs}

\end{table*}

\begin{figure*}
    \includegraphics[width=\textwidth, angle=0]{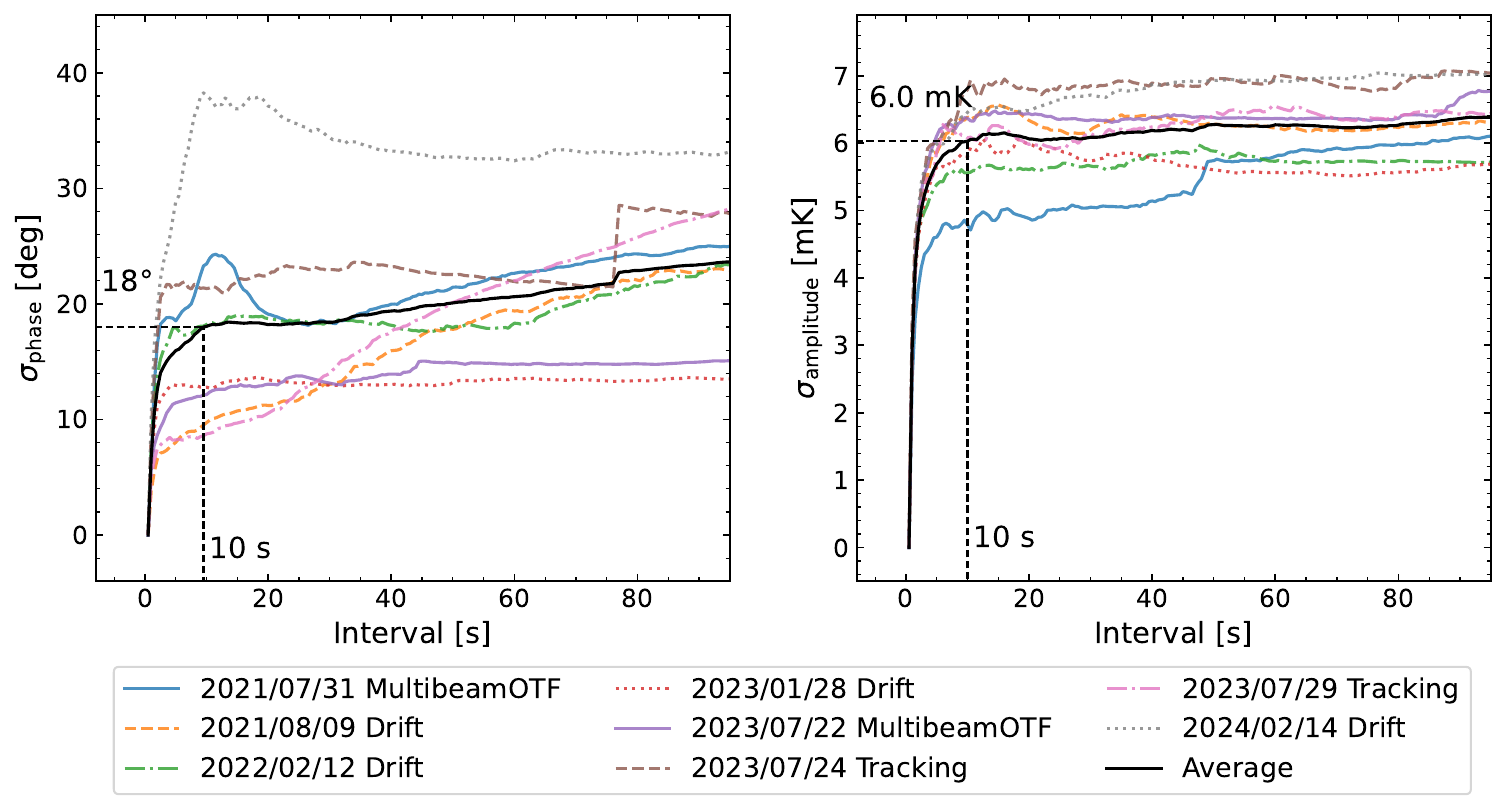} 
    \caption{Left panel: The standard deviation of phases in different time intervals. We use eight days' observations by MultibeamOTF, Tracking, or drift-scan modes. The black solid curve is the average of eight days. 
    For example, it is easy to read out that the phases change 18\degree\ in 10 seconds on average.   
    Right panel: Similar to the left, but for the amplitude. The amplitudes change to 6 mK in just 10 seconds, as shown with the black dashed lines. 
    }
	\label{fig:sig}
\end{figure*}

The left and right panels of Figure~\ref{fig:sig} present the median value of the variance in phases and amplitudes $\sigma$ for eight observations individually over various time intervals. The frequency range from 1390 MHz to 1440 MHz is used here. Solid black curves represent the average value of eight observations. 
The standard deviation ($\sigma$) of both phase and amplitude increases significantly as time intervals increase. After 10 seconds, the standard deviation of the phase will reach 18 degrees and the standard deviation of the amplitude will reach 6 milliKelvin. 
These results imply that rapid changes occur in both phase and amplitude in a short time frame, potentially stemming from the variations in systematic temperature and gain (or zenith angle), the shape of the active reflector, and other intricate factors.
This suggests that traditional standing wave removal methods, such as the running average method over long intervals, may not be effective for quickly changing data. A removal process based on individual spectra is likely to yield better results.

\subsection{Standing Waves in the Noise Diode Modulated Mode}\label{sec:noise}

\begin{figure*}
    \includegraphics[width=0.5\textwidth, angle=0]{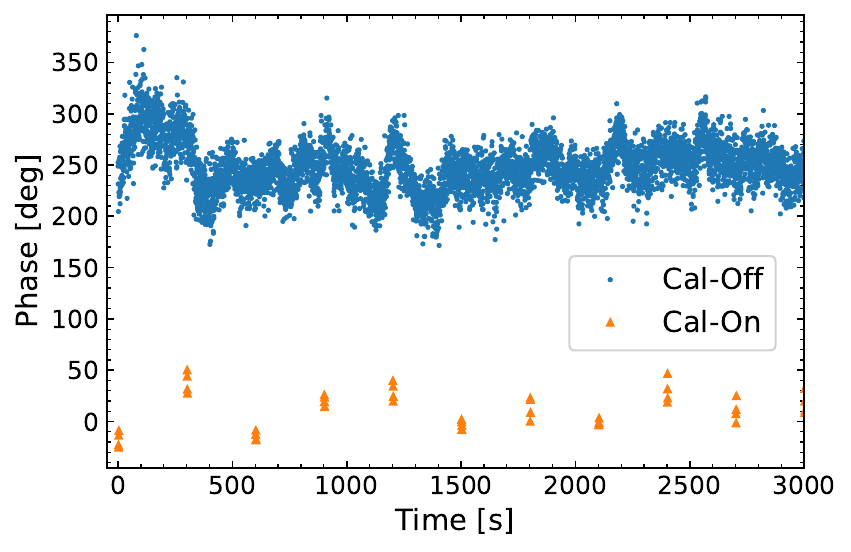}\centering 
    \includegraphics[width=0.5\textwidth, angle=0]{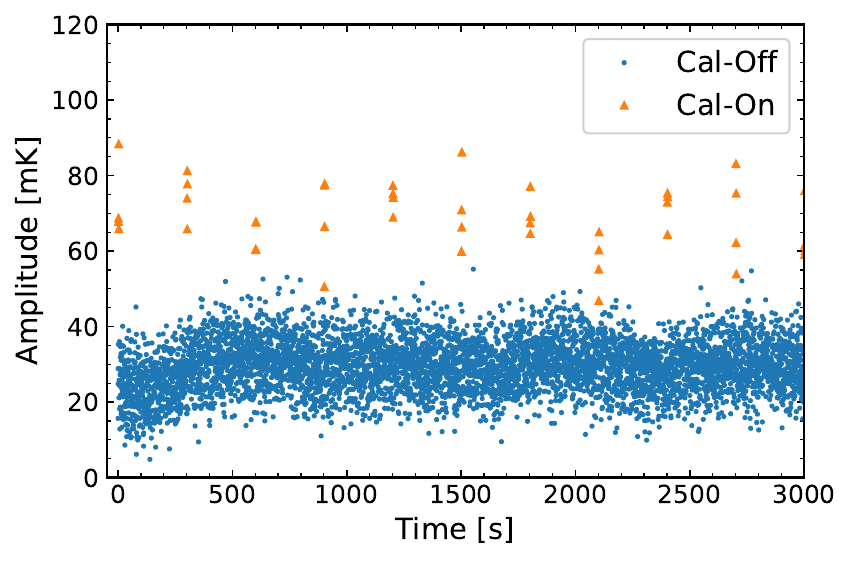}\centering
    \caption{The phases (left panel) and amplitudes (right panel) of the 1-MHz standing wave in the drift mode. The orange triangles indicate that when the noise diode is on, the phase shifts by approximately 200\degree\, and the amplitude increases by 4 times compared with the Cal-Off spectra (blue dots).
    }
    \label{fig:noise_diode}
\end{figure*}

The noise diode is utilized to calibrate the intensity to the antenna temperature ($T_\mathrm{A}$) in each observation, as described in detail in the \HIFAST\ paper. However, the noise diode device may modulate the properties of the standing wave. We investigate the behaviour of the standing wave in noise-diode-modulated mode. 
Typically, during most of our observations, the noise diode remains active for 2 seconds and then switches off for another 298 seconds, with a delay of 2 seconds at the observation beginning. Figure~\ref{fig:noise_diode} illustrates the phase and amplitude of each spectrum in a 300-second observation. 
The results with the noise diode on (Cal-On) or off (Cal-Off) are indicated individually by orange and blue symbols.

The period of the 1-MHz standing wave during the switch of these two modes does not change significantly, but the phase and amplitude exhibit significant changes when the noise diode is turned on: the phase shifts by approximately 200\degree\, and the amplitude increases by a factor of 4. 
This implies that the spectrum with the diode on should be treated separately. Moreover, the stronger standing wave when the diode is on results in a slightly larger RMS of the entire spectrum (systematic temperature) compared to when the diode is off.

\section{Standing Wave Removal Methods}\label{sec:methods} 
There are typically a variety of approaches for removing standing waves, such as sine-fitting, moving median, and FFT filtering. This section will briefly outline these techniques individually and subsequently assess their effectiveness using FAST data in relation to the characteristics discussed in Section \ref{sec:features}.

\subsection{Sine-fitting: Traditional Sine Function Fitting}\label{sec:sin}
As the standing wave resembles a sinusoidal curve, it is common practice to remove it using a sine-fitting function. Typically, the sinusoidal function utilized to fit the standing wave can be expressed as
\begin{equation}
    T_{\mathrm{SW}}(\nu) = A_0 + A_1\sin(2\pi \tau \nu + \phi),
    \label{equ:sin}
\end{equation}
where $A_0$ is the constant displacement and $\nu$ corresponds to the frequency. 
Here $\tau$ is $1 / f$, the time delay of the standing wave. $\phi$ and $A_1$ are the phase and amplitude of the standing wave, respectively.

To reduce the number of unknown fitting parameters, \citet{Peek2011, Peek2017} conducted a fitting procedure on the temperature differences using two components: a series of sine (and cosine) functions and a Taylor expansion of the residual \HI contribution. Conversely, \citet{Nidever2010} opted to keep the amplitude and wavelength constant while allowing the phase to vary. 
For this research, we initialized a time delay of $\tau = 0.92$ \mus, as determined from Equation~\ref{equ:sdw-1} for the standing wave of 1 MHz. 
Subsequently, the wave was modeled using a sine function through iterations of the least-squares algorithm. This approach is similar to the technique employed by \citet{Zheng2020}, where they fitted a sinusoidal wave with a fixed period derived from the FFT transfer of the spectrum.

\subsection{Running Median: Reference Baseline by Moving Window} \label{sec:running}

If the bandpass, baselines, and standing waves are gradually variable, a widely used technique, running median, could be applied to subtract the continuous components in the spectra by computing the median value in the moving window along the time axis. 
When this method removes the continuum background, we should be cautious about whether it could reserve the signal features. 
Therefore, the size of the window should be carefully determined.
A larger window size means that the signal will lose less flux, decreasing the bias by a strong source like \citet[Fig. 3]{Barnes2001}. 
The window size should also be limited because we cannot estimate standing waves from a long timescale over which the phases and amplitudes may have changed significantly, as depicted in Section~\ref{sec:amp_phi}. 
Considering a typical point source whose diameter is 7 arcmin, it requires 28 seconds to scan across with the drift-scan mode.
As discussed above, we use a moving window of 300 s integration to calculate the median as the reference baseline and mask the source to avoid overfitting when comparing it with other methods. 

\subsection{FFT Filter: Sinusoidal Modes Filter by FFT}

\begin{figure*}
    \includegraphics[width=\textwidth, angle=0]{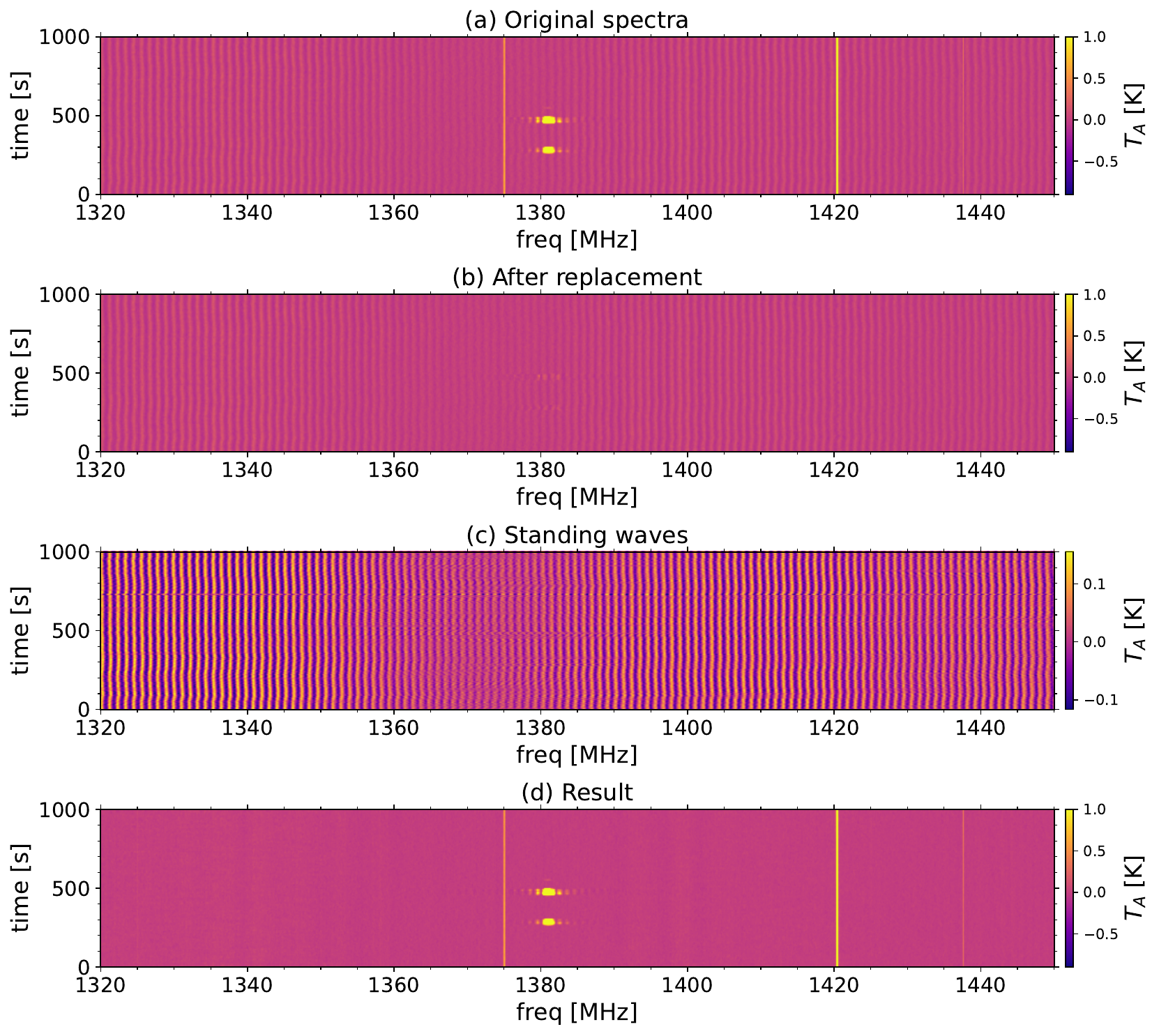}\centering
    \caption{An example of the procedure of the FFT filter. Panel (a) is the original input data after the baseline subtraction. After the replacement of RFI and Milky Way (MW), strong signals are filled with standing waves as panel (b). Then the components of the standing waves are chosen from the Fourier space and presented in the spectral space as panel (c). Finally, panel (d) is panel (a) minus panel (c), which is the result after the extraction of standing waves. The colour corresponds to the antenna temperature in Kelvin. A Gaussian interpolation is applied helping to display the ripples.}
    \label{fig:waterplot}
\end{figure*}

It is intuitive to subtract the standing wave using the Fourier transform. 
A significant method was presented by \citet{Briggs1997}, who adjusted the FFT algorithm and experimented with it on the Arecibo and Nançay telescopes. Subsequently, \citet{Barnes2005} revised the technique to process the Parkes multibeam survey data. 
They modelled the amplitudes, deployed a phase-tracking algorithm for the harmonic modes, and directly eliminated the standing waves from the data. 
In this research, we apply an enhanced approach that does not involve fitting in Fourier space. Figure~\ref{fig:waterplot} provides an example with 1000s, 130 MHz spectra to demonstrate the FFT filtering technique.

\subsubsection{Preprocess and replacement}\label{sec:rep}

\begin{figure}
    \includegraphics[width=0.5\textwidth]{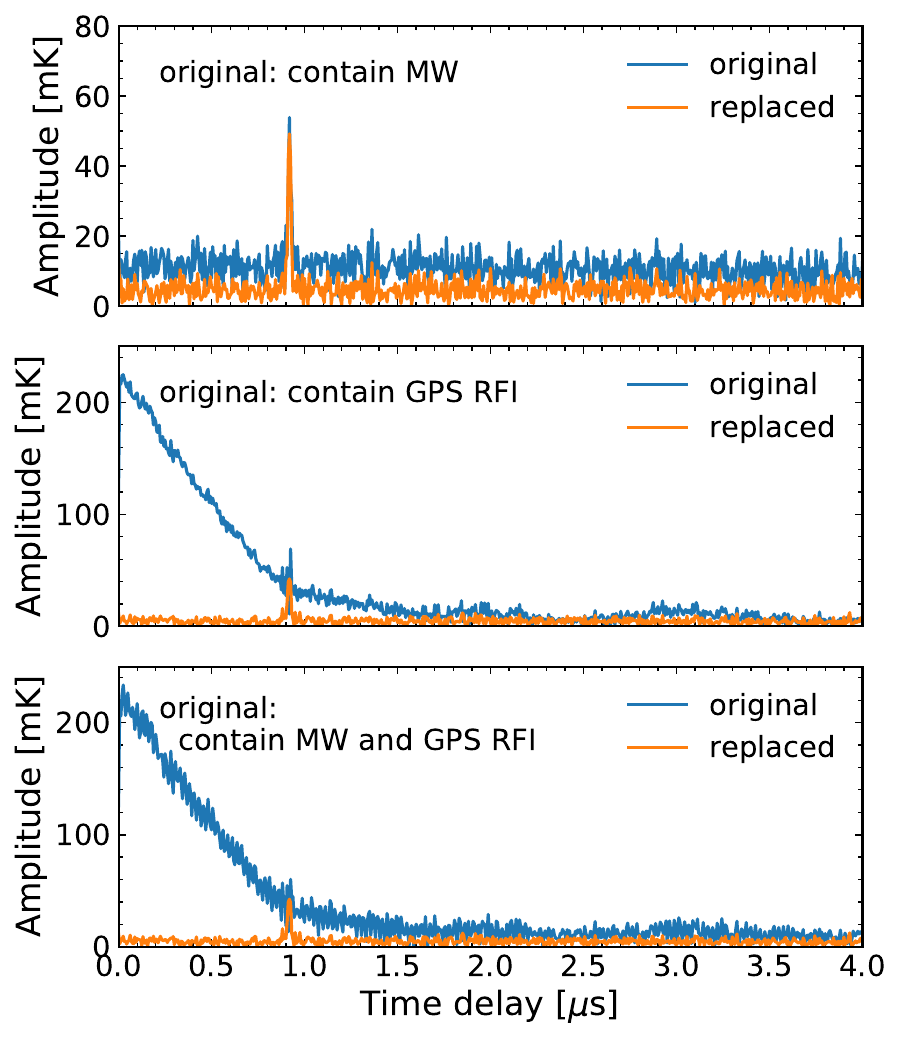}\centering
    \caption{The Fourier amplitudes of spectra with MW, GPS RFI, and both, from top to bottom, respectively. The peak of the 1-MHz standing wave is blended with noise from MW and a strong background from RFI in the original (blue) spectrum, while it could be simply separated from the `replaced' (orange) spectrum.}
    \label{fig:rep_reason}
\end{figure}

The Fourier modes of standing waves mix with various complex components, potentially originating from RFI, the Milky Way (MW), continuum, or unidentified sources. Prior to performing the Fourier transform, it is essential to preprocess the data by subtracting the baseline and masking the RFI regions to prevent the introduction of additional Fourier modes.
If the \HI signals and RFI are not masked, the standing wave modes will be considerably mixed with other components, as illustrated in Figure~\ref{fig:rep_reason}.
The blue spectrum is generated from the spectrum with only the baseline subtracted, in which the MW or GPS RFI\footnote{Global Positioning System L3 beacon at 1380 -- 1382 MHz} remain original.
The first panel of Figure~\ref{fig:rep_reason} shows that extended signals like MW bring a flat background and noise for the blue spectrum.
In the second panel, RFI and other emissions contribute to strong modes at the lower end of time delay. 
Considering the situation with both MW and RFI in the bottom panel, Fourier modes of the original spectrum are severely polluted, increasing the difficulty of separating the 1-MHz mode.
While for the orange spectrum Fourier transformed from the `replaced' spectrum, all RFI and strong signals are cleared, and it exhibits an apparent peak in 0.92 \mus\ of the 1-MHz standing wave.  
It demonstrates the necessity to replace RFI and strong sources before the FFT process.

To mitigate the effects of interference and emissions, some works employed a curve fitting procedure to eliminate any bright lines, although this method is intricate and time-intensive. Therefore, in this research, we adopt a straightforward assumption that the standing waves are roughly identical within close frequency ranges. 
Consequently, for convenience, we replace RFI and signals exceeding the threshold with frequency-adjacent standing waves. 
We note that only narrow-band RFI that affects only 1 -- 2 channels (see \citetalias{Jing2024}) is set to noise. 

\begin{figure}
    \includegraphics[width=0.5\textwidth]{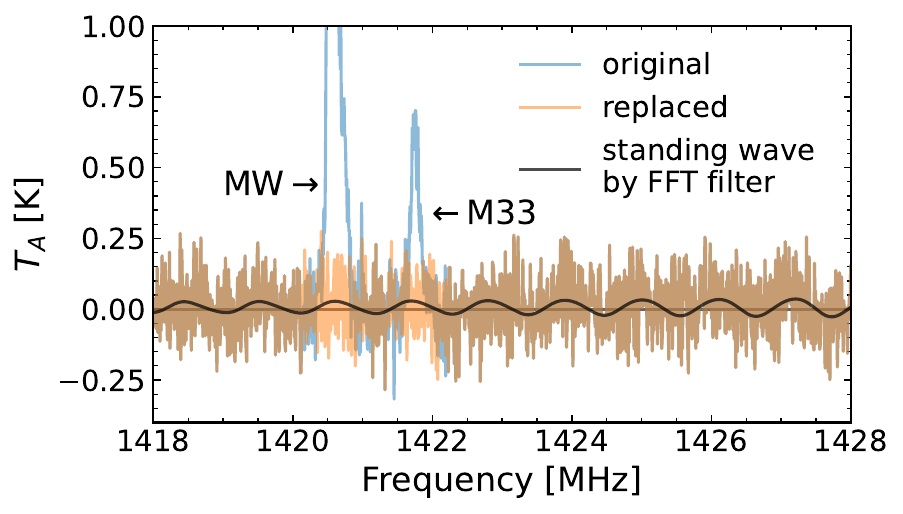}\centering
    \caption{Illustration of the replacement in a spectrum containing MW and M33. The MW and M33 (blue peaks in the spectrum) are located at 1420 -- 1422 MHz. We process FFT on the `replaced' (orange) spectrum and extract the black standing waves that fit exactly. In this way, we can estimate the flux hidden in the extended signals. 
 }
    \label{fig:rep_m33}
\end{figure}

After the preprocessing, the modes of the standing waves have become quite apparent, as can be seen from Figure~\ref{fig:rep_reason}.
To illustrate this process, we display a spectrum in the MW and M33 (Triangulum galaxy) area in Figure~\ref{fig:rep_m33}. 
Substituting the adjacent standing waves allows us to approximate the behaviour of standing waves in the presence of strong extended sources. The results demonstrate that the shape of the ripple remains ideal after replacement.

Thus, we obtain the processed data without significant sources and RFI, as depicted in Figure~\ref{fig:waterplot}. 
Considering our interest in flux, evaluating the consequences of the replacement is crucial. 
This methodology is believed to minimally affect the phases of standing waves because the replaced segment spans only 1 -- 2 MHz, which is much narrower compared to the frequency range applied for FFT in our analysis. In cases of substantial RFI such as GPS interference, the replacement segments can extend to approximately 5 -- 20 MHz. 
Simply zero-padding would worsen the spectral leakage due to truncation, resulting in less efficient fitting around the masked segments. Therefore, we avoid setting zeros directly in these regions.

\subsubsection{Identification of standing wave modes}
Next, we could process the FFT on the two-dimensional data that have replaced RFI and signals. For each spectrum, the quasi-monochromatic standing wave contains several modes during the discrete transform,
\begin{equation}
    T_{\mathrm{SW}}(\nu) = A_0 + \sum_{i=1}^{n} A_i\sin(2\pi \tau_i \nu + \phi_i),
    \label{equ:fft}
\end{equation}
where $n$ is the number of modes that contribute to the standing waves. Here, $A_0$ is still a constant value. $A_i$, $\phi_i$, and $\tau_i$ are the amplitude, phase, and time delay of the $i^{\rm{th}}$ mode.
We categorize the types of standing waves by taking the average of the amplitudes and utilizing the peak location to determine the characteristic time delay of the standing wave. The FFT is executed along the frequency axis, and as a result, each spectrum develops its own set of Fourier modes based on the individual processed spectrum.

\subsubsection{Removal of standing wave modes}
\begin{figure}
    \includegraphics[width=0.5\textwidth]{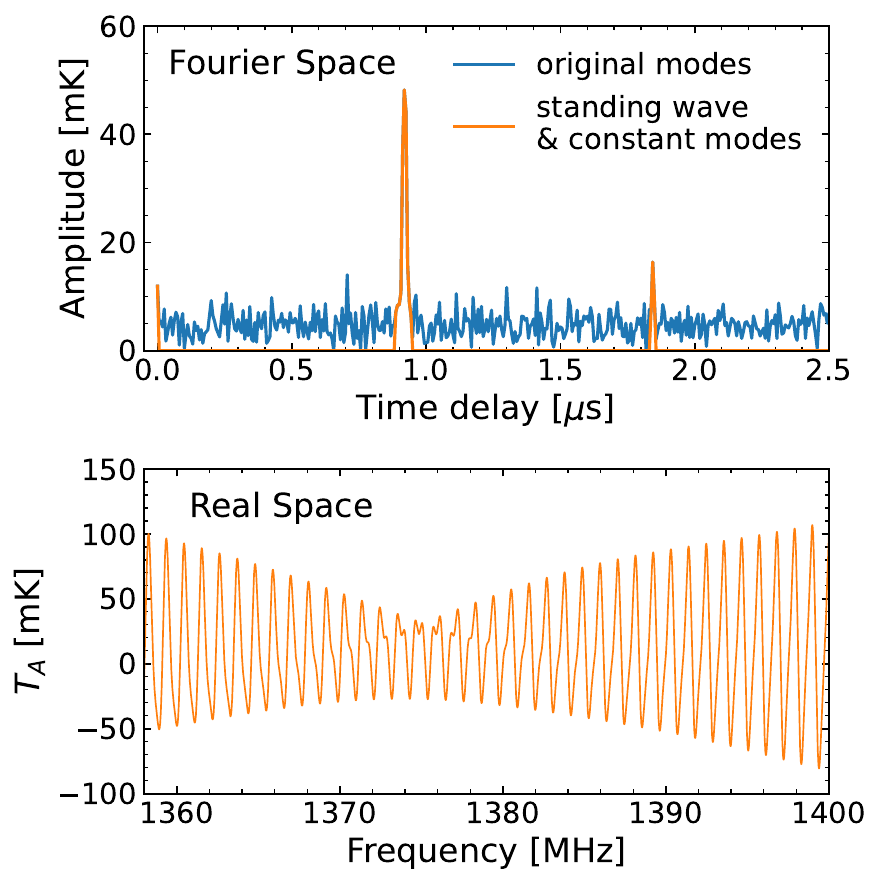}\centering
    \caption{Upper panel: Amplitudes of one `replaced' spectrum (blue) in Fourier space. We chose the orange components of the 1-MHz standing wave and the constant mode to perform the iFFT. 
    Lower panel: Standing wave packet from the iFFT. Then we could effectively subtract the standing waves in real space.}
    \label{fig:choose}
\end{figure}

After acquiring the phases and amplitudes of the spectra via the FFT method, our subsequent step is to isolate the components associated with the standing waves. For each spectrum in the dataset, we select the modes that meet two specific criteria to carry out the inverse FFT (iFFT):
\begin{enumerate}[(1)]
    \item  The amplitude of modes should surpass a specified threshold.
    \item  These modes are situated within a specific time-delay range around the identified standing waves.
\end{enumerate}
For instance, the orange modes illustrated in the upper panel of Figure~\ref{fig:choose} satisfy these criteria, contributing to the 1-MHz standing wave. 
Because of sampling and spectral leakage effects, multiple modes were selected to encompass the entire peak. 
Furthermore, a constant ``zero" component can be removed to reduce baseline deviation. 
The chosen modes are then transformed into a wave packet in real space, as shown in the lower panel of Figure~\ref{fig:choose}. 
Consequently, we achieve the standing wave with a period of 1.09 MHz along with its harmonic waves. 
Panel (c) of Figure~\ref{fig:waterplot} displays the standing waves determined through this process.

~\\
In conclusion, as illustrated in Figure~\ref{fig:waterplot}(d), we eliminate the standing waves derived from the iFFT, resulting in data free of ripples. Additionally, standing waves can be isolated from spectra that have not undergone baseline correction to enhance the outcome of the subsequent baseline subtraction.

\section{Implications in the Real Data}\label{sec:result}

To assess the effectiveness of removing standing waves, we begin by evaluating the reduction in RMS after applying the FFT filter method. As previously mentioned, we calculate the theoretical RMS ($\sigma_{\mathrm{theory}}$ in Equation~\ref{equ:theory_rms}) of the averaged spectrum. 
During a five-minute integration, the FFT filter method reduces the RMS from 3.2 $\sigma_{\mathrm{theory}}$ to about 1.15 $\sigma_{\mathrm{theory}}$. 
For longer integration, standing waves contribute more to the RMS than Poisson noise, making this decreasing trend more pronounced and highlighting the importance of standing wave removal.

~\\
In the following sections, we will present and summarize the implications of different standing wave removal methods applied to the real observational data.

\subsection{Standing Waves of Three Methods}

\begin{figure*}
    \includegraphics[width=\textwidth, angle=0]{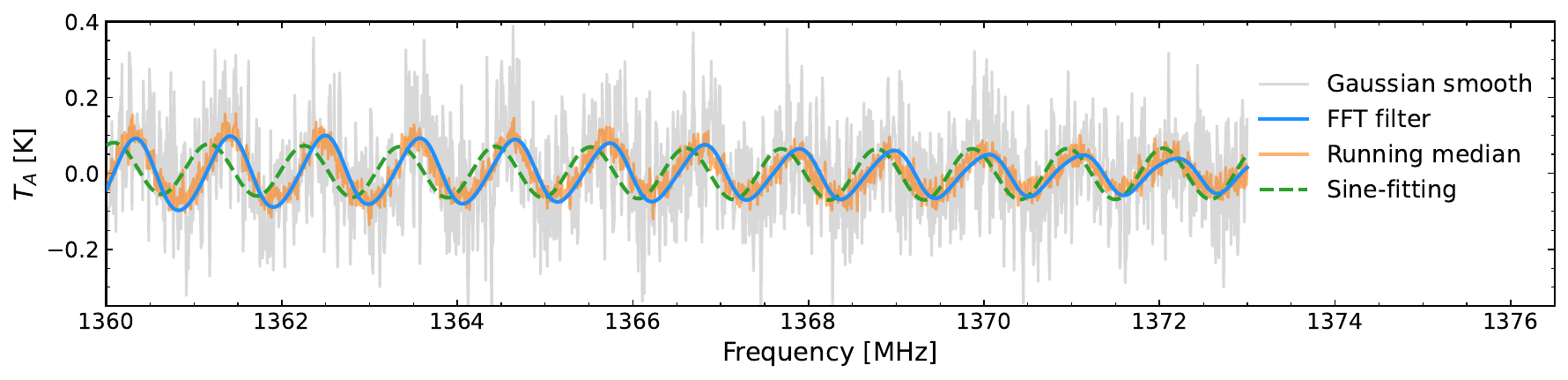}\centering
    \caption{Standing waves obtained from different methods: FFT filter (blue solid line), running median (orange solid line), and sine-fitting (green dashed line). The grey spectrum in the background is Gaussian smoothed from the signal-free area. 
    Here we average the spectra over 5 seconds to increase the signal-noise rate (SNR). }
    \label{fig:sw_eff}
\end{figure*}

We analyse the standing waves obtained through three different methods: sine-fitting, running median, and FFT filter. 
Figure~\ref{fig:sw_eff} displays the standing waves over a signal-free section of the spectrum in real space. 
The standing waves produced by sine-fitting do not align accurately with the grey Gaussian-smoothed spectrum at the side of 1360 MHz, due to the complexity of standing wave fitting. 
In contrast, the running median and the FFT filter method both provide acceptable results for standing waves compared to Gaussian smoothing.
An obvious difference is that the standing wave derived from the FFT filter exhibits a smooth profile, while the standing wave obtained by the running median retains more noise.
Additionally, the FFT filter and running median methods can simultaneously remove standing waves of 1-MHz, 1.92-MHz, and 0.039-MHz. 
It is simpler to eliminate several modes of standing waves in Fourier space rather than fitting sine functions repeatedly.

\subsection{RMS Estimation of Point Sources} \label{sec:point_sources}

Alongside the qualitative comparison, we also quantitatively evaluate the noise level after eliminating the standing wave for point sources.
In this study, 26 point sources free from serious RFI were identified in a two-day drift scan using positions from the Arecibo Legacy Fast ALFA survey catalogue \citep[ALFALFA, ][]{Haynes2011, Haynes2018}. 
We measure the RMS from three data cubes separately, which follow the same data reduction routine except for the standing wave removal procedure.
When determining the source's total flux density, we need to integrate the pixels over the aperture area (see Eq. 3 of \citet{Haynes2018}), so
the theoretical RMS estimated here is proportional to $\sqrt{N_\mathrm{pix}}$ or $\sqrt{t_\mathrm{int}}$, the pixel number or integration time of each source. Therefore, the aperture to calculate the source spectrum has been controlled the same for comparison. 

\begin{figure}
    \includegraphics[width=0.5\textwidth]{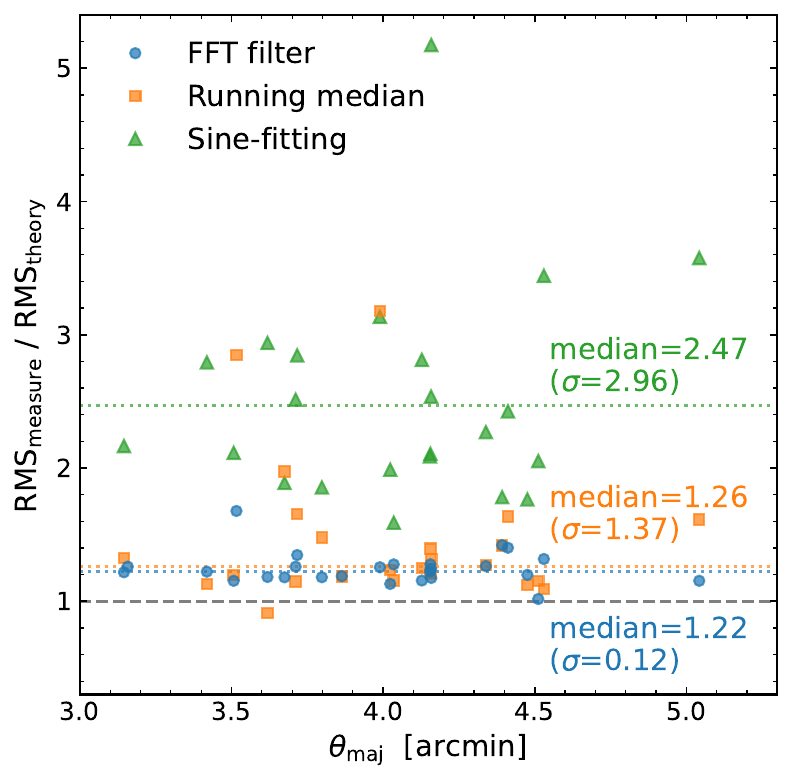}\centering
    
    \caption{The RMS ratio between measurement and theoretical estimation using three standing wave removal methods: the FFT filter (blue circles), running median (orange squares), and sine-fitting (green triangles). 
    $\theta_{\mathrm{maj}}$ is measured as the elliptical major axis of the source. 
    A black dashed line is plotted for the reference equal to the theoretical noise level.
    Here we measured 26 point sources in the FAST data, but some points of the sine-fitting method exceed the upper limit of this figure.  
    The coloured dotted lines correspond to the median value of the FFT filter, running median, and sine-fitting, and the $\sigma$ represents the standard deviation of each method.
    The FFT filter presents the lowest median RMS of $1.22~\sigma_\mathrm{theory}$ and the smallest scatter of 12\%.}
    \label{fig:rms_ratio}
\end{figure}

Figure~\ref{fig:rms_ratio} shows the measured RMS to theoretical RMS ratio among three standing wave methods.  
The parameter $\theta_{\mathrm{maj}}$ denotes the major axis size of the ellipse for each point source as measured from the FFT method products, corresponding to the convolved size of the source and beam pattern, so all the sources are point sources with size larger than 3 arcminutes. 
The coloured dotted lines indicate the median values of the RMS ratios labelled on the right, where its scatter is denoted in the bracket beneath the corresponding lines.

Most of the data points derived from the sine-fitting technique are values significantly greater than one, whose median has reached $2.47~\sigma_\mathrm{theory}$.
The sine-fitting also exhibits the largest scatter, signifying a poor spectral quality when assessing RMS.

The median RMS of running median is slightly greater than the ideal theoretical expectation, being $1.26~\sigma_\mathrm{theory}$.
This method can remove anomalies such as RFI, signals, or baseline fluctuations to some extent, which may explain why it displays the noise level quite close or even lower than ideal estimation in some cases. 
However, since the standing wave is not always stable in the moving window we selected, the varying phases and amplitudes may cause the misalignment when subtracting the standing waves using running median, resulting in some sources with RMS $> 1.5 ~\sigma_\mathrm{theory}$ and a scatter of 137\%.
As mentioned in Section~\ref{sec:running}, using spatial filters on extended astronomical sources involves a significant risk of losing flux.
A common example is the negative bandpass bias for highly extended and strong sources \citep{Barnes2001} unless additional correction.
Here we apply a source mask based on the prior knowledge from ALFALFA to reserve the flux, whereas for \HI\ blind surveys, the FFT filter is more appropriate.

In Figure~\ref{fig:rms_ratio}, the FFT filter provides the closest median RMS to the theoretical value ($1.22~\sigma_\mathrm{theory}$), which reaches the noise level of the running median.
Furthermore, the FFT filter performs the smallest scatter, 12\% of the derived spectrum noise, indicating its universal applicability for the fluctuating standing waves of FAST.
The FFT filter has also preserved the source flux through the replacement and the two-step iteration procedure, and we will describe it in the last discussion.

\section{Harmonic RFI: Other Harmonic Components during FFT}\label{sec:rfi}

\begin{figure*}
    \includegraphics[width=\textwidth, angle=0]{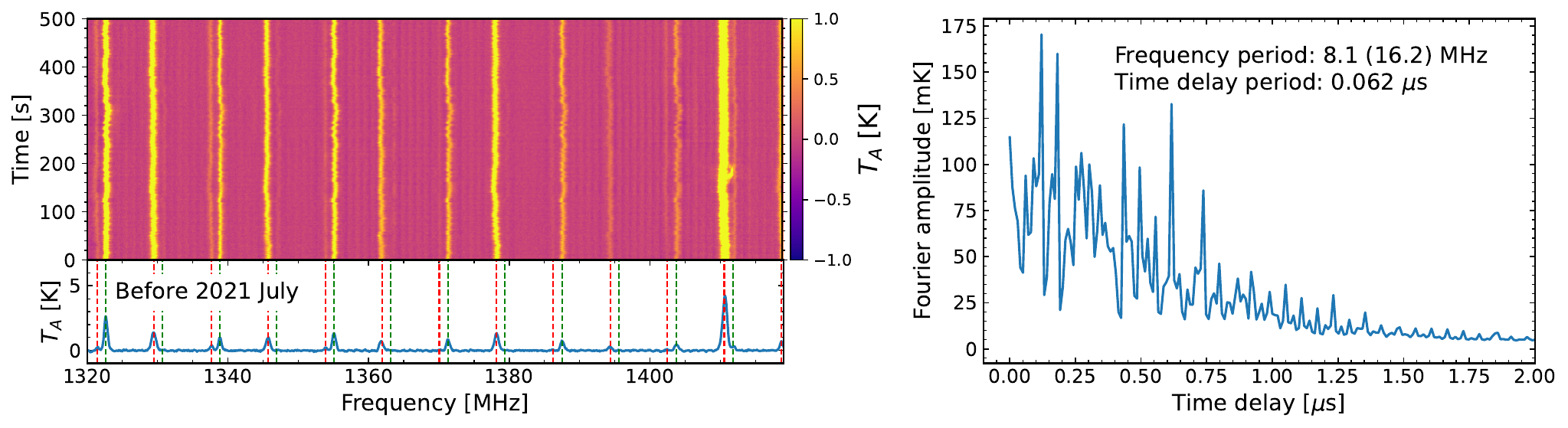}\centering
    
    \includegraphics[width=\textwidth, angle=0]{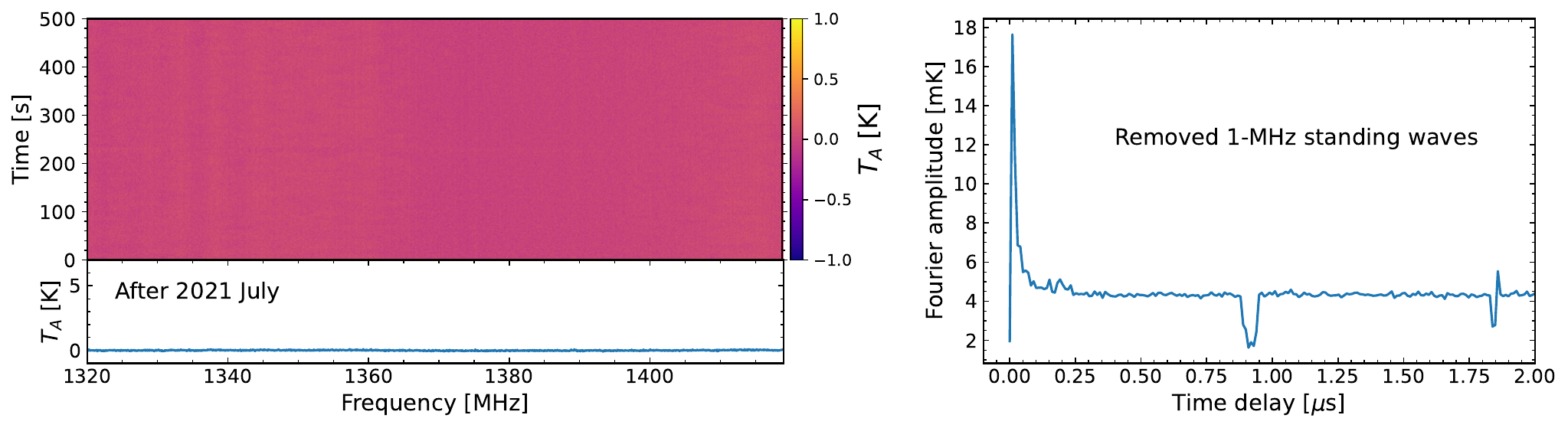}\centering
    
    \includegraphics[width=\textwidth, angle=0]{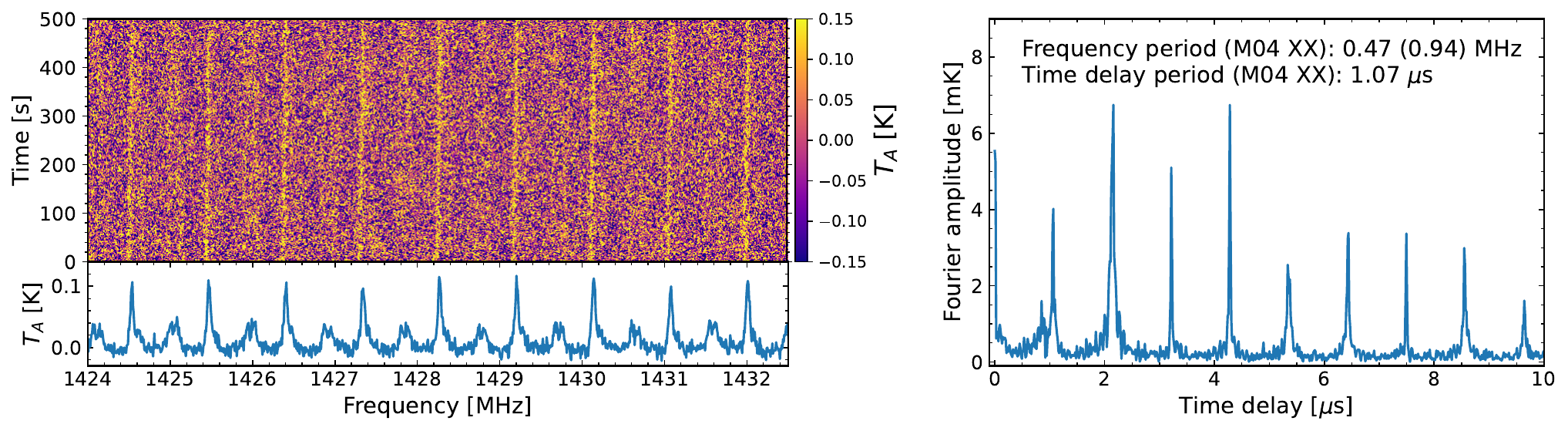}\centering
    
    \includegraphics[width=\textwidth, angle=0]{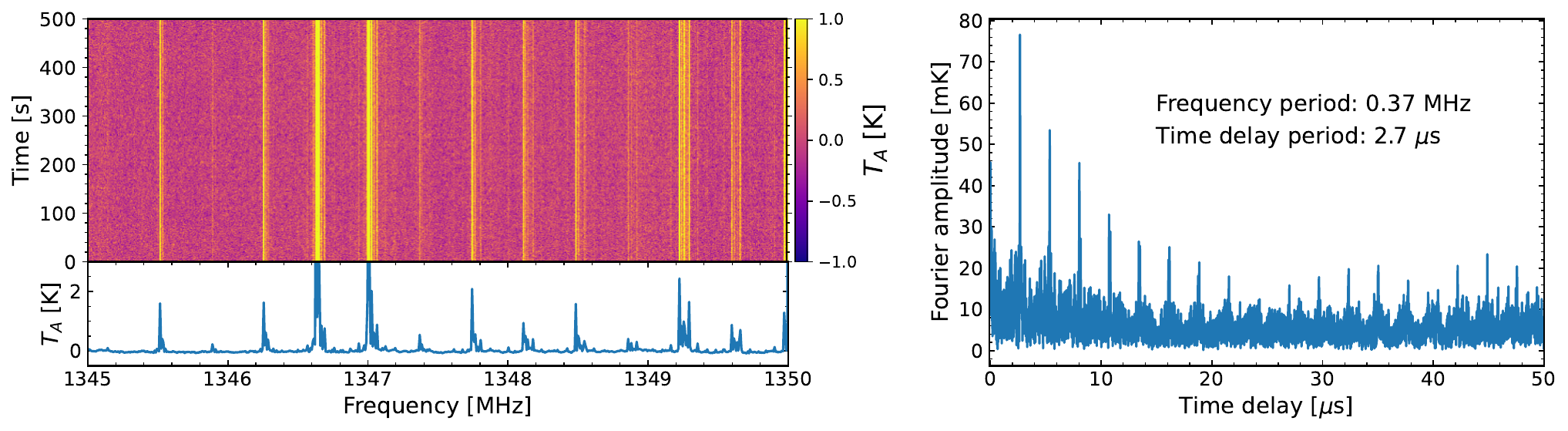}\centering
    
    \caption{Harmonic RFI in FAST spectrum data. The left panels depict the RFI in real space. The colour scale corresponds to the antenna temperature measured in Kelvin. Below the waterfall plot, the spectrum is smoothed or averaged along the time axis. The right panels represent the Fourier-transformed version of the left panels, revealing the harmonic features inherent in the RFI.
    \textbf{8.1-MHz RFI:} The first row shows the 8.1-MHz RFI before July 2021. Two sets of Gaussian-like peaks are highlighted with dashed lines in red and green. Each set has a period of nearly 8.1 MHz, but only the strong peaks are obvious, leading to a characteristic period of 0.062 \mus\ (that is 1/16.2 MHz) in the right panel. 
    After July 2021, the 8.1-MHz RFI has been successfully eliminated, as shown in the second row. The 1-MHz standing waves have been also removed in those panels. Refer to Section~\ref{sec:8rfi} for more details.
    \textbf{0.5-MHz RFI:} The third row illustrates the 0.5-MHz RFI. In the waterfall plot, equally spaced pulses are apparent with a period close to 1 MHz. Other smaller peaks become discernible in the spectra below if we average more spectra. As this period is related to specific beams and polarisations, we use M04 \texttt{XX} for example. The details of its accurate period are described in Section~\ref{sec:0.5rfi}.
    \textbf{0.37-MHz RFI:} The last row exhibits the 0.37-MHz RFI, characterized by a period of 0.37 MHz in frequency and 2.7 \mus\ in time delay. The intensity of this RFI varies from several Kelvins to hundreds, so we display only a portion of it in real space. Further elaboration can be found in Section \ref{sec:0.37rfi}.
    }
    \label{fig:pdrfi}
\end{figure*}

The FFT is a valuable tool for examining standing waves and can also be utilized to analyse harmonic RFI. While both harmonic RFI and standing waves show harmonic properties, the former is more complex in the Fourier domain. 
The features of these RFI signals in Fourier space are harmonic modes, allowing us to easily identify the harmonic components of RFI and determine their period. 
The familiar period in MHz is the reciprocal of the period in time-delay space. 
For FAST, we categorized three types of harmonic RFI based on their frequency periods: 8.1 MHz, 0.5 MHz, and 0.37 MHz. 
Table~\ref{tab:rfi} summarizes their main characteristics. 
The intensity, in Kelvin, indicates the peak temperature in real space. 
To assess the impact of RFI, we use the RMS over a two-minute integration and a 5-MHz bandwidth, as described previously. Figure~\ref{fig:pdrfi} shows their performance in both real space (left columns) and Fourier space (right columns).

\begin{table*}[ht]
	\setlength{\tabcolsep}{6pt}
	\renewcommand\arraystretch{1}
	\caption{Basic information and current status of the harmonic RFI at FAST. 
	}
    \begin{center}
	\begin{tabular}{cccccc}
		\hline\noalign{\smallskip}
		Period / MHz & Time delay / \mus & Intensity / K & RMS$^*$ / $\sigma_{\mathrm{theory}}$ & Eliminated time & Comments\\
		\noalign{\smallskip}
        \hline
        \noalign{\smallskip}
	  8.1 & 0.062 & \s~ 1 -- 10 & 11.1 & after 2021/07 & $^a$ \\
	0.5 & 1 & \s~ 0.01 -- 0.1 & 1.2 &  - & $^b$ \\
	  0.37 & 2.7 & mostly $<$10 & 23.0 & - & $^c$ \\
		\noalign{\smallskip}\hline
	\end{tabular}      
    \end{center}
{\small
$^*$ RMS in 5-MHz and 2-minute integration.

$^a$ The 8.1-MHz RFI from compressors has been eradicated at the end of July 2021. Its time delay of 0.062 \mus\ corresponds to 1/16.2 MHz, consistent with the prominent RFI peaks. Here the two polarisations are merged.

$^b$ We have known the origin of the 0.5-MHz RFI and are trying to reduce those pulses. The time-delay period of the dominated RFI series is \s~1 \mus. Each polarisation of 19 beams has a fixed but diverse period, so we note that the approximate period of all of them is \s~0.5 MHz in real space. 
The RMS contribution of 0.5-MHz RFI is measured using M04, \texttt{XX} only for instance.

$^c$ This RFI is more prevalent on the low-frequency side of the Ultra-Wide Bandwidth (UWB) receiver at FAST. It only manifests in the $L$-band 19-beam receiver for a few days at the end of November 2022. Here we refer to M08, \texttt{XX} for example.
}
\label{tab:rfi}
\end{table*}

\subsection{The 8.1-MHz RFI} \label{sec:8rfi}

In the data archived before July 2021, the 8.1-MHz harmonic RFI was found consistently in all 19 beams and polarisations, creating considerable challenges for data processing. \cite{Jiang2020} described it as the 1-MHz wide RFI during that time. 
The bright stripes in the first row of Figure~\ref{fig:pdrfi} represent the peaks of the harmonic RFI, which show a slight drift along the time axis during observations. 
When examining the smoothed spectrum below the waterfall plot, the Gaussian-shaped peaks can be divided into two or three distinct clusters, indicated by red and green dashed lines. As mentioned earlier, this RFI comes from three compressors located in the feed cabin, which are used for cooling the receiver. 
This likely explains why the RFI appears in three distinct sets. Occasionally, the third set may not be easily seen or it may overlap with other features, causing it to be missing in the spectra. 
In Fourier space, the RFI modes exhibit equally spaced characteristics with a period of 0.062 \mus, equivalent to the reciprocal of twice the 8.1-MHz frequency. Due to the alternating strength of the RFI peaks, the true period is more likely to be 16.2 MHz.

By late July 2021, shielding was applied to the compressors, causing the 8.1-MHz harmonic RFI to vanish, which is illustrated in the second row of Figure~\ref{fig:pdrfi}. 
The Fourier amplitude spectrum now shows an almost flat behaviour, except for the residual 1-MHz standing waves seen after prolonged integration. 
Unlike standing waves, harmonic RFI cannot be eliminated due to the presence of too many equally spaced modes. Reconstructing the harmonic RFI by employing a series of sine components with different amplitudes is unfeasible. This key principle holds true for other RFIs, like the 0.5-MHz and 0.37-MHz RFIs discussed later. 
To reduce the influence of the 8.1-MHz RFI in previous data, \HIFAST\ determines the exact period from the peak positions in the spectrum and then masks the polluted frequency regions in real space.

\subsection{The 0.5-MHz RFI} \label{sec:0.5rfi}

The third row of Figure~\ref{fig:pdrfi} shows some weak pulses that are evenly spaced with a period close to 1 MHz, especially visible in the waterfall plot. The underlying spectrum highlights the complex nature of this RFI due to its intricate configuration.
To examine this RFI, we continued utilizing the FFT, and the right panel illustrates the distinct modes with a separation of nearly 1 \mus. 
Importantly, the characteristic period differs for each beam and polarisation. For M04 \texttt{XX}, the precise period is 0.94 MHz (equivalent to 1.07 \mus).
However, we detected a series of very weak peaks positioned between two strong peaks separated by 1 MHz in real space. By using a similar approach as for the 8.1-MHz RFI, we verified that the true period of this 1-MHz RFI is roughly 0.5 MHz for 19 beams, and for M04 \texttt{XX}, it is 0.47 MHz. 
Moreover, the presence of two sets of 0.5-MHz pulses with a separation of approximately 0.1 MHz adds to the complexity of the averaged spectrum. 
Occasionally, only a single set of pulses with their prominent peaks is noticeable, prompting others to call it the 1-MHz RFI sometimes.

Currently, we ignore the dense pulses, as they are only noticeable in certain polarisations like M08 \texttt{YY} and M04 \texttt{XX} among others. 
However, the harmful effect of this harmonic RFI on the baselines during extended integration periods is still recognized. 
For a two-minute integration, the 1-MHz peaks with an intensity of \s~0.05 K remain undetectable, since the RMS only increases to 1.2 times the theoretical RMS. On the other hand, for a 30-minute integration, the RMS exceeds 2.3 $\sigma_{\mathrm{theory}}$. 
Luckily, we have identified the source of this RFI --- the Analog-to-Digital Converter (ADC) in the digital backends. This information could help eliminate the interference in the future.

\subsection{The 0.37-MHz RFI} \label{sec:0.37rfi}

The 0.37-MHz RFI is predominantly found on the low-frequency end (below 1 GHz) of FAST's latest Ultra-Wide Bandwidth (UWB) receiver, which spans 0.5 -- 3 GHz \citep{Zhang2023}. 
The strength of the dense pulses varies from a few Kelvins to hundreds, significantly impacting observations on the high-redshift side. 
Notably, the 0.37-MHz RFI was detected during three days of observations using the $L$-band 19-beam receiver, as illustrated in the last row of Figure~\ref{fig:pdrfi}. 
Next, we convert the RFI into Fourier space, revealing a series of peaks with a 2.7 \mus\ period that corresponds to 0.37 MHz. 
Upon closer inspection of a single pulse in real space, additional sub-structures with an approximate period of 0.013 MHz become evident, although these are not shown in Figure~\ref{fig:pdrfi}. 
We deduce that the UWB receiver generated the 0.37-MHz RFI, inadvertently affecting the $L$-band 19-beam receiver on select days in late November 2022.

\section{Conclusions and Discussion} \label{sec:conclusion}

FAST and some other radio telescopes face similar issues with standing waves. However, the standing waves seen at FAST are particularly complex and unstable due to the unique active main reflector. 
A major problem for FAST is the presence of a 1-MHz standing wave, which needs to be addressed. To tackle this, we have carried out a comparative study of commonly used techniques for standing wave elimination, such as sine-fitting, running median, and FFT filter. 
Below, we summarize our findings on the characteristics of the standing waves and the efficiency of these techniques.

\begin{enumerate}[(1)]
    \item FAST's standing waves are classified into three types: 1.09, 1.92, and 0.039 MHz. Following the instrumentation enhancements in 2022, the 0.039-MHz standing wave and the second harmonic of the 1-MHz standing wave have been diminished (see Table~\ref{tab:sw}).
    
    \item The phases and amplitudes of the standing waves exhibit an apparent variability. Specifically, the standard deviation of phases can extend up to 18 degrees, while the amplitudes undergo a change of 6 mK within a ten-second interval (see Fig.~\ref{fig:sig}). 
    Consequently, a prolonged integration period to create a reference spectrum for standing wave removal is infeasible for FAST.
    
    \item Noise diode also affects the standing waves, leading to a sudden increase in amplitude and an obvious phase drift when injecting noise (see Fig.~\ref{fig:noise_diode}).

    \item The FFT filter technique effectively reduces the RMS of the spectrum from 3.2 times the theoretical RMS to 1.15 times, bringing it closer to the anticipated sensitivity level for a five-minute integration.

    \item Alternative approaches such as sine-fitting and running median cannot effectively eliminate standing waves for all the cases. 
    According to the RMS estimation from 26 point sources, the FFT filter presents the lowest median RMS, also \s~1.2 times the theoretical expectation.
    Besides, it displays the smallest scatter of 12\%, showing the best stability compared with the other two methods (refer to Fig.~\ref{fig:rms_ratio}).
    
\end{enumerate}

~\\
In this research, we have evaluated three methods for mitigating the standing waves in the FAST's observations, focusing on their intuitive visual impact and their performance based on the RMS estimation.
And now our focus has shifted to examining how these methods affect the measurement of source flux.

Figure~\ref{fig:sw_eff} reveals that the misaligned standing wave obtained from the sine-fitting introduces spurious flux, necessitating a more cautious control of initial parameters especially when handling large datasets. 
In addition, the signals from \HI sources, particularly in cases of faint double-horned profiles, may inadvertently blend with a sine function, leading to overfitting and ultimately the underestimation of the flux.

To prevent the flux loss problem, the running median method needs to choose a sufficient moving window compared to the source scale. 
However, it is ineffective since the phases of the standing waves would have varied significantly. 
This change subsequently influences the position of the fitted standing wave and, consequently, the large RMS deviation to theoretical prediction. 
The SoFiA manual also stresses that the median filter should be used cautiously as it may impact emissions from galaxies. 
\citet{Barnes2001} and \citet{Putman2002} investigated MINMED5 or MEDMED5, which offer alternative methods to avoid negative bandpass sidelobes during the median process \citep[e.g.,][]{Putman2003, Minchin2010}. These techniques are used for bandpass subtraction in \HIFAST.

The FFT filter, removing the stand waves for every spectrum, remains unaffected by the time-varying phases and amplitudes. 
As for the flux problem, although we applied a replacement procedure for RFI and strong sources before the Fourier filter, there exists a chance that a faint source might not meet the threshold for substitution. In this case, we adopt a two-step approach: remove the standing waves in the first iteration and use the intermediate result to redefine the area exceeding the threshold in the second iteration. 
This approach effectively replaces most of the signals, particularly for sources with a lower SNR concealed in the standing waves. 
Importantly, since the Fourier modes we select account for only a small fraction of the spectrum, the risk of excessive filtering remains relatively low. 

Additionally, we know that FFT is apparently useful for standing wave analysis. It is also suitable for harmonic RFI examinations and period recognition. 
So we confirm three types of harmonic RFI we found at FAST, whose periods are 8.1, 0.5, and 0.37 MHz, respectively (refer to Table~\ref{tab:rfi} and Fig.~\ref{fig:pdrfi}).

In general, the FFT filter stands out as the most effective method for mitigating time-varying standing waves of FAST.
It is compatible with both extended sources and point sources, unless the data length is insufficient, leaving a low resolution in Fourier space. 
In most cases, the FFT filter performs much better than traditional running median/mean and sine-fitting methods, preserving flux without introducing spurious signals.
Despite instrumental efforts to weaken standing waves, their interference with radio spectra persists over time.
The FFT filter could solve the problem, which makes up the standing wave removal section of \HIFAST, and some research has been based on our programs \citep[e.g.,][]{Zhang2024, Liu2024a, Pan2024}. 
In the future, observations of FAST can rely on these procedures to improve the spectrum quality, ultimately showcasing the telescope’s high sensitivity using data products with standing waves mitigated.


\begin{acknowledgements}
We would like to thank the referee for the constructive suggestions and comments. 
We also thank Yinghui Zheng and Hengxing Pan for their insightful discussions on FAST's data processing.
This work was supported by the China National Key Program for Science and Technology Research and Development of China (2022YFA1602901, 2023YFA1608204), the National SKA Program of China (No. 2022SKA0110201), the National Natural Science Foundation of China (Nos. 11873051, 11988101, 12033008, 12041305, 12125302, 12173016, 12203065), the CAS Project for Young Scientists in Basic Research grant (No. YSBR-062), 
the K.C. Wong Education Foundation, and the science research grants from the China Manned Space Project. 
Y.J. acknowledges support from the Cultivation Project for FAST Scientific Payoff and Research Achievement of CAMS-CAS.

This work has used the data from the Five-hundred-meter Aperture Spherical radio Telescope (FAST). FAST is a Chinese national mega-science facility, operated by the National Astronomical Observatories of the Chinese Academy of Sciences (NAOC).
\end{acknowledgements}


\bibliographystyle{raa}
\bibliography{ms2024-0293}

\label{lastpage}

\end{document}